# ARTICLE

# The Meissner effect in a strongly underdoped cuprate above its critical temperature

Elvezio Morenzoni[1], Bastian M. Wojek[1,2], Andreas Suter[1], Thomas Prokscha[1], Gennady Logvenov[3,†] & Ivan Božović[3]

The Meissner effect and associated perfect 'bulk' diamagnetism together with zero resistance and gap opening are characteristic features of the superconducting state. In the pseudogap state of cuprates, unusual diamagnetic signals and anomalous proximity effects have been detected, but a Meissner effect has never been observed. Here we probe the local diamagnetic response in the normal state of an underdoped $La_{1.94}Sr_{0.06}CuO_4$ layer ($T_c' \lesssim 5\,K$), which is brought into close contact with two nearly optimally doped $La_{1.84}Sr_{0.16}CuO_4$ layers ($T_c \approx 32\,K$). We show that the entire 'barrier' layer of thickness, much larger than the typical $c$ axis coherence lengths of cuprates, exhibits a Meissner effect at temperatures above $T_c'$ but below $T_c$. The temperature dependence of the effective penetration depth and superfluid density in different layers indicates that superfluidity with long-range phase coherence is induced in the underdoped layer by the proximity to optimally doped layers, but this induced order is sensitive to thermal excitation.

[1] Laboratory for Muon Spin Spectroscopy, Paul Scherrer Institute, 5232 Villigen PSI, Switzerland. [2] Physik-Institut, Universität Zürich, 8057 Zürich, Switzerland. [3] Condensed Matter Physics & Materials Science Department, Brookhaven National Laboratory, Upton, New York 11973-5000, USA. †Present address: Technology Service Group, Max-Planck Institut für Festkörperforschung, 70569 Stuttgart, Germany. Correspondence and requests for materials should be addressed to E.M. (email: elvezio.morenzoni@psi.ch).

I n cuprates that show high-temperature superconductivity (HTS), the 'normal' (metallic) state above the critical temperature $T_c$ is quite anomalous, featuring linear temperature dependence of resistivity, a pseudogap in the density of states, absorption throughout the infrared region, signatures of local superconducting correlations and so on[1–6]. Surprisingly, the superconducting state seems to be quite ordinary in many respects. One unusual feature is the so-called Giant Proximity Effect, reported by several groups[7,8]. In one such experiment, trilayer ('sandwich') Josephson junctions were fabricated with top and bottom electrodes made of S=$La_{1.85}Sr_{0.15}CuO_4$ with $T_c≈45$ K, and barrier layers made of S'=$La_2CuO_{4+δ}$ with a 'typical' $T_c'≈25$ K (ref. 8). Such SS'S junctions were shown to transmit supercurrents even at temperature $T > T_c'$ and to do so for barriers much thicker (up to 20 nm) than expected from the standard theory of the proximity effect[9,10]. The problem is that Josephson tunnelling experiments are delicate. Great effort was applied to exclude contributions from filamentary superconductivity, but its absence could only be concluded from circumstantial evidence[8].

One would prefer a direct measurement that can detect superconductivity, access buried layers, probe locally, scan along the $z$ coordinate and estimate the superconducting volume fraction in each layer, thus differentiating between 'filamentary' and 'bulk' superconductivity. The local probe technique we have chosen for this study (low-energy (LE) muon spin rotation (μSR) in fact fulfils all these requirements. By directly mapping the magnetic field profile in the heterostructure, we show that the entire underdoped (UD) barrier layer displays superconductivity with an effective critical temperature lower than, but close to, that of the optimally doped (OP) top and bottom layers.

## Results

**Magnetic field profiling.** LE-μSR has previously been applied to obtain the depth-resolved profile of local magnetization in various thin films and heterostructures on the nm scale[11,12]. By tuning the muon implantation energy, we vary the stopping range of the fully spin-polarized muons between 3 and 200 nm, and measure timedifferential μSR spectra at different depths in the heterostructure (Fig. 1). The spectra yield information about the magnetic and superconducting properties in different layers. For instance, the observed muon spin precession frequency is directly proportional to the local magnetic field at the muon site and can detect a diamagnetic shift associated with supercurrents. In a magnetic field with a broad field distribution, one observes fast damping directly related to the width of the distribution. The technique is presented in the Methods section.

**Film growth and characterization.** The films were synthesized in a molecular beam epitaxy system designed for atomic-layer engineering of complex oxide materials. The atomic scale smoothness and high quality of cuprate heterostructures grown in this equipment has enabled to realise one-unit-cell-thick HTS or insulating layers[13] and interface superconductivity in metal-insulator bilayers[14]. For the present study, we fabricated a number of heterostructures, as well as single-phase films for control experiments. The results are shown for heterostructures consisting of three layers, each 46-nm thick; OP $La_{1.84}Sr_{0.16}CuO_4$ ($T_c≈32$ K) was used for the top and the bottom 'electrodes', whereas UD $La_{1.94}Sr_{0.06}CuO_4$ ($T_c'≲5$ K) served as the 'barrier'. The barrier with a low $T_c'$ offers a broad temperature interval to search for putative long-range proximity effects. Similar results have been obtained with 32 nm thick barriers.

The films were grown simultaneously on four single-crystal $LaSrAlO_4$ substrates, each with the $10×10$ mm² surface polished parallel to the (001) plane, under nominally identical conditions (in $1.2·10^{-5}$ mbar of ozone and at the substrate temperature of about 700 °C) and annealed in high vacuum to avoid inadvertent oxygen doping. The deposition rates were measured by a quartz crystal oscillator before growth and controlled in real time using a custom-made atomic absorption spectroscopy system[15]. The quality of the film growth was checked by monitoring reflection high-energy electron diffraction (RHEED) intensity oscillations, which provide digital information on the film thickness. Subsequent ex situ characterization included mutual inductance, X-ray diffraction (XRD), resistivity, atomic force microscopy (AFM) and Rutherford backscattering measurements. Mutual inductance was measured for every film, individually, in the transmission geometry, that is, with the film placed in-between the drive and the pick-up coils. A lock-in amplifier was used at the measurement frequency $ν = 10$ kHz. Comparison of mutual inductance data with four-terminal resistivity measurements shows that, as the temperature is decreased, the resistance vanishes at the critical temperature $T_c$, which exactly corresponds to the temperature below which the inductive component starts to decrease and the dissipative component starts to increase.

To enhance the signal-to-noise ratio, the μSR measurements of the trilayers were done simultaneously on a set of three nominally identical films (3 cm² total area). For the single layers (SLs), the set consisted of four films (4 cm² total area). Figure 2a shows that all the trilayers have sharp superconducting transitions with $T_c≈32$ K with very small variations in $T_c$ within the set of three trilayer films. In contrast, all single-phase $La_{1.94}Sr_{0.06}CuO_4$ films show $T_c'≲5$ K. Here, the variations in $T_c$ within the set are somewhat larger; they are probably due to small variations in the Sr doping level or in the density of oxygen vacancies, and the greater sensitivity of $T_c$ to such variations for doping in the vicinity of the superconductor–insulator transition. Note, however, that this does not affect our conclusion in the paper. Figure 2b and c shows reflectivity and XRD, respectively, from a $La_{1.84}Sr_{0.16}CuO_4/La_{1.94}Sr_{0.06}CuO_4/La_{1.84}Sr_{0.16}CuO_4$ trilayer measured at room temperature on a Seifert XRD 3003 PTS with a 4 Ge 220 crystal monochromator, providing Cu $K_{α1}$ radiation with wavelength 0.15406 nm. The XRD patterns show very high crystallinity and absence of secondary phases. Rocking curves with full-width at half-maximum smaller than 0.015° and finite-thickness oscillations visible in low angle reflectivity curves and around the $(002n)$ Bragg peaks ($n = 1,2,3$) indicate the high quality of the films and interfaces. From the finite thickness oscillations the overall film thickness has been determined to be $136±1$ nm, consistent with the values determined by Rutherford backscattering (Fig. 2d). The root mean square (r.m.s.) roughness of the interfaces as obtained from the fit of the X-ray reflectivity curve is about 1 nm. The typical surface roughness determined by AFM was 0.5 nm, much less than one-unit-cell height (1.3 nm).

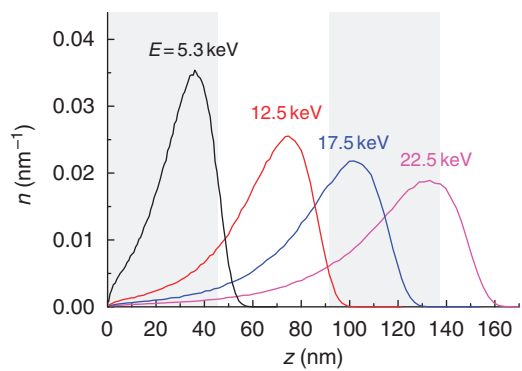

**Figure 1 | Stopping profiles.** Normalized muon implantation profiles in the trilayer structure $La_{1.84}Sr_{0.16}CuO_4$ (46 nm)/$La_{1.94}Sr_{0.06}CuO_4$ (46 nm)/ $La_{1.84}Sr_{0.16}CuO_4$ (46 nm) at some of the energies used. The grey shaded areas indicate the top and bottom layers of the heterostructure.

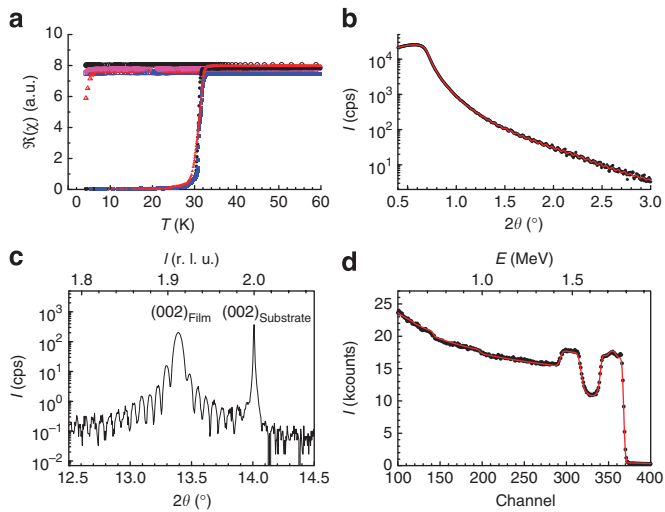

**Figure 2 | Sample characterization. (a)** The inductive response of three $La_{1.84}Sr_{0.16}CuO_4/La_{1.94}Sr_{0.06}CuO_4/La_{1.84}Sr_{0.16}CuO_4$ trilayers (filled symbols) and four single-phase $La_{1.94}Sr_{0.06}CuO_4$ films (open symbols). Each set was deposited simultaneously and under nominally identical conditions. **(b)** X-ray reflectivity of a trilayer film. The Kiessig fringes reflecting the finite film thickness indicate atomically smooth surface and interfaces. The continuous curve is a fit to the data obtained with the GenX routine[33]. **(c)** XRD data of a trilayer film showing the finite-thickness oscillations in the vicinity of the (002) Bragg reflection. **(d)** Typical Rutherford backscattering spectrum of a trilayer film. The curve is a simulation obtained by the RUMP program[34] including multiple scattering effects and gives a total thickness of 140 nm (±4 nm). The rms surface roughness as measured by atomic force microscopy is smaller than 0.5 nm over an area of $2,500\,\mu m^2$.

**Magnetic properties.** We first discuss the magnetic properties of the UD layer grown as a SL or as a barrier in the trilayer structure (TL). The comparison of the magnetic behaviour in the two cases, which strongly depends on the doping level, allows us to assess the equivalence and integrity of the layers. For this we implant in the centre of the layer muons with 5.3 and 12.5 keV, respectively, and record muon spin precession/relaxation spectra at zero field (ZF) and in a field $B_{ext} = 9.5$ mT applied parallel to the ab planes and transverse to the initial muon spin polarization (TF). Typical ZF $\mu SR$ asymmetry spectra, which are proportional to the muon polarization, $A(t) = A_0 P(t)$, are shown in Figure 3. The time evolution of the polarization of the muons implanted in the layer can be consistently fitted in both cases by a combination of a fast relaxing and a slowly relaxing component. (In the top and bottom OP layers only the component weakly relaxing due to the small nuclear moments is present, reflecting the absence of magnetism in these layers.)

The temperature dependence of the ZF asymmetry data was fitted with the CERN $\chi^2$ minimization routine MINUIT2 in the following way:

$$A(t) = A_0 P(t) = (A_s e^{-\lambda_s t} + A_f e^{-\lambda_f t}) e^{-\frac{\sigma_{nucl}^2 t^2}{2}}$$
$$+ A_{Ni} e^{-\lambda_{Ni} t} + A_{tb} e^{-\frac{\sigma^2 t^2}{2}} \quad (1)$$

The first term in parentheses describes the dominant signal from the UD layer of the samples (SL or TL), which were mounted on a pure Al plate covered with Ni. The signal from the layer consists of a fast (f) and a slowly (s) relaxing component. The fast component describes the 'spin-glass' like state. The slowly relaxing component describes the non-magnetic or weakly magnetic volume fraction of the layer.

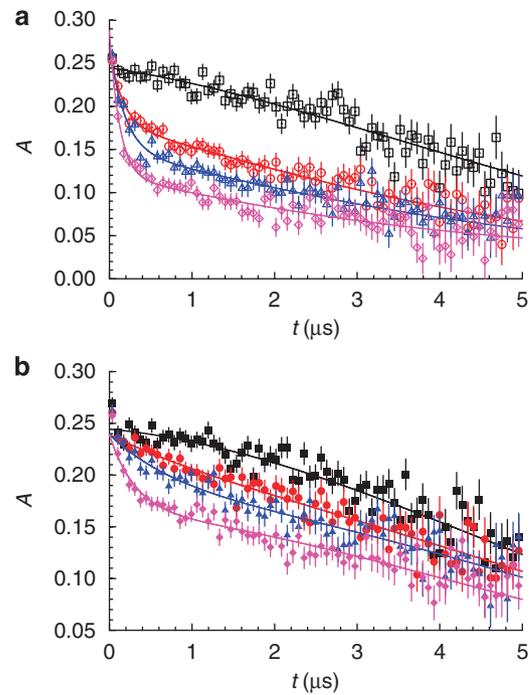

**Figure 3 | $\mu SR$ spectra in ZF. (a)** ZF asymmetry spectra from the single UD layer (black squares $T = 40$ K , red circles 8.3 K, blue triangles 6.3 K, magenta diamonds 4.6 K). **(b)** Spectra from the trilayer OP/UD/OP structure (black squares 40 K, red circles 8.0 K, blue triangles 6.0 K, magenta diamonds 4.4 K). The energy is chosen in both cases to stop the muons at the centre of the UD layer. The pronounced depolarization at low temperatures and its evolution is similar in both cases. The higher baseline in the trilayer case reflects essentially the –17% weakly relaxing contribution of the top layer. Error bars indicate the statistical uncertainty of the data, curves are fits to the data (see text for details).

The sum of the two contributions $A_f + A_s$ is temperature independent and constant. The spin depolarization caused by the nuclear dipolar fields has been determined at 200 K and fixed to $\sigma_{nucl} = 0.14\,\mu s^{-1}$. It is multiplied with the electron moment contribution because it is uncorrelated with it. The role of Ni is to very quickly depolarize the muons that miss the sample and therefore to suppress their contribution to the measured signal. $A_{Ni}$ and $\lambda_{Ni} = 88\,\mu s^{-1}$ have been determined at 40 K and kept constant at lower temperature. $A_{tb}$ corresponds to muons stopping in the top or the bottom layer of the heterostructure. At 12.5 keV the corresponding fraction is 17% of the total sample signal with a damping rate in the top layer $\sigma = 0.14\,\mu s^{-1}$ independent of temperature. In the SL case $A_{tb} = 0$.

To analyse the temperature scans in the transverse field case, the $A_s$ and $A_{tb}$ terms in equation (1) are multiplied by an oscillating term. Figure 4 shows the temperature dependence of the amplitude of $A_f$ and $A_s$ and their relaxation rates ($\lambda_f$ and $\lambda_s$) in this case. The relaxation rate is a measure of the width of the distribution of local magnetic fields ($\lambda \simeq \gamma_\mu \Delta B$), which is proportional to the size of the magnetic moments, but also may include contributions from fluctuating fields. The emergence of $\lambda_f$ below ~15 K is due to the build up of random static fields from Cu electronic moments. The fact that we observe two distinct $\mu SR$ signals in TF and ZF means that we have two spatially separated phases. A fraction $A_f/(A_f + A_s)$ of the muons probes a magnetic phase, whereas the others experience only very weak magnetic fields. The non-observation of a spontaneous spin precession shows that the magnetic phase originates from the moments that slow down and freeze with random orientation producing a distribution of random static fields. In La$_{2-x}$Sr$_x$CuO$_4$ this short-range antiferromagnetic correlated spin-glass-like state (sometimes termed cluster spin-glass) is known to persist at $x > 0.02$ coexisting with superconductivity in the strongly UD regime $0.05 < x < 0.1$ (ref. 16). The magnetic structure is heterogeneous on a length scale larger than a few nm, as the fields produced by the copper magnetic moments decay away over this length scale. The evolution with temperature of the signals shows that, while the total amplitude remains constant, the magnetic phase begins to appear below about 15 K and gradually grows at the expense of the other as the temperature decreases (see Fig. 4). Unlike $A_f(T)$, $\lambda_f$ is almost temperature independent. This indicates that the disorder and size of the moments immediately saturate on freezing, whereas an increasing volume fraction of the UD layer becomes magnetic as the temperature is lowered further.

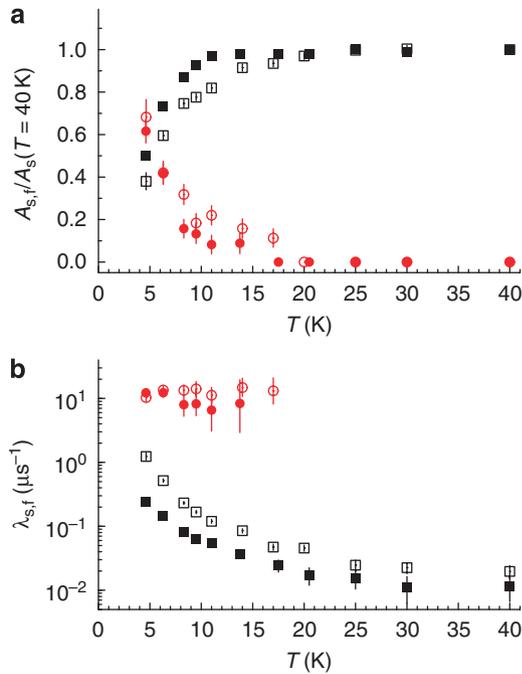

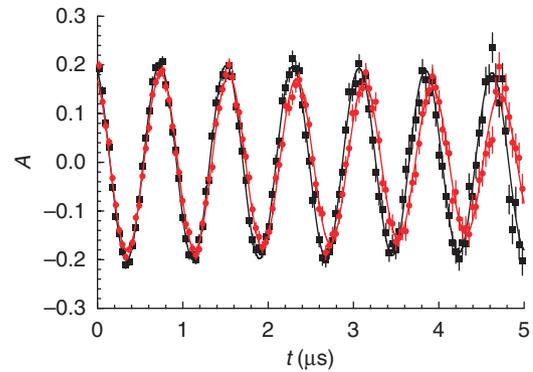

**Figure 5 | Muon spin rotation signal.** Time dependence of the muon decay asymmetry (proportional to the polarization) for muons implanted at 12.5 keV in the central underdoped layer ($T_c' \lesssim 5$ K) of the trilayer structure La$_{1.84}$Sr$_{0.16}$CuO$_4$ (46 nm)/La$_{1.94}$Sr$_{0.06}$CuO$_4$ (46 nm)/La$_{1.84}$Sr$_{0.16}$CuO$_4$ (46 nm). An external magnetic field of 9.5 mT was applied parallel to the $ab$ planes and to the interface (Meissner geometry). The signal at 40 K (black) represents the muon spin precession in the external field. At 9.5 K (red), well above $T_c'$, the lower precession frequency (proportional to the average local field) reflects the presence of diamagnetism in the barrier. Error bars indicate the statistical errors of the data, curves are fits to the data (see text for details).

**Figure 4 | Signal amplitudes and relaxation rates.** (a) Temperature dependence of the relative fractions of fast (f, circles) and slowly (s, squares) relaxing components for a 46-nm thick underdoped La$_{1.94}$Sr$_{0.06}$CuO$_4$ layer as a single-phase film ($E = 5.3$ keV, open symbols) or as barrier in a trilayer heterostructure ($E = 12.5$ keV, closed symbols). (b) Corresponding relaxation rates (proportional to the width of the field distribution) for both configurations. Note the increase in amplitude of the fast relaxing component below ~15 K indicating the development of disordered magnetism. Error bars give the fit errors.

Important for the following considerations about the diamagnetic response is that the very similar temperature dependence of the parameters characterizing the magnetic phase of the SL and TL case confirms the integrity of the barrier and its position in the La$_{2-x}$Sr$_x$CuO$_4$ phase diagram.

**Diamagnetic response.** To map the diamagnetic response of the heterostructure as a function of position along the crystal axis ($z$ coordinate), we cooled the samples in ZF from above $T_c$ to ~4.3 K, applied a magnetic field of 9.5 mT parallel to the $ab$ planes ($x$ direction) and collected $\mu$SR spectra as a function of the muon implan-tation energy at increasing temperatures. The weakly damped spin precession observed at an angular frequency $\omega_L$ corresponding to a mean local field $\langle B_x \rangle \omega_L / \gamma_\mu$ (where $\gamma_\mu$ muon gyromagnetic ratio) exhibits the diamagnetic response of the heterostructure. This is clearly seen in Figure 5, which demonstrates the main result of this paper: at a temperature well above $T_c'$, the spins of the muons implanted in the barrier coherently precess in a field that is diamagnetically shifted with respect to the applied field. As the maximum diamagnetic shift is small (0.2–0.3 mT) for an applied field of 9.5 mT, and the experimentally measured field distribution resulting from weighting the profile $B_x(z)$ with the implantation profile can be well approximated by a Gaussian distribution, the asymmetry term can be written as $\propto e^{-\sigma^2 t^2/2} \cos(\gamma_\mu \langle B_x \rangle t + \phi)$. $\langle B_x \rangle$ is obtained from a fit of the asymmetry spectra over a time interval (for $t > 0.15$–0.5 $\mu$s) where the fast relaxing parts give negligible contribution.

The depth profile of the mean field $\langle B_x \rangle$ at different temperatures is shown in Figure 6. At 10, 15 and 17 K—that is, well above $T_c'$—the local field is lower than the applied field at all depths, meaning that the entire heterostructure excludes the magnetic flux like a conventional superconductor. This is unexpected when one recalls that in this geometry the supercurrent must pass through the 'barrier' La$_{1.94}$Sr$_{0.06}$CuO$_4$ region that is 46-nm thick. This is over two orders of magnitude larger than the $c$ axis coherence length $\xi_c$ in the electrodes. Note that at $T > T_c'$ a single-phase La$_{1.94}$Sr$_{0.06}$CuO$_4$ is not superconducting, and not even metallic along the $c$ axis. In Figure 7a, we compare the temperature dependence of the average field in the centre of a single-phase film of UD La$_{1.94}$Sr$_{0.06}$CuO$_4$ with that in the barrier of the same composition inside a trilayer heterostructure. In the former case no shift is observed, whereas in the latter case the shift is observable up to $T_{di} \approx 22$ K. The observed field profile reflects the shielding supercurrent that runs along the $c$ axis as well as in the $ab$ planes of the barrier; note that $\langle j_{ab} \rangle = \langle (1/\mu_0) dB_x(z)/dz \rangle \neq 0$. The profile has the form of an exponential field decay in the Meissner state with the flux penetrating from both sides and looks like that for two superconductors with different magnetic penetration depths.

We modelled the field by a solution of the London equations and obtained the magnetic penetration depths in the electrodes,

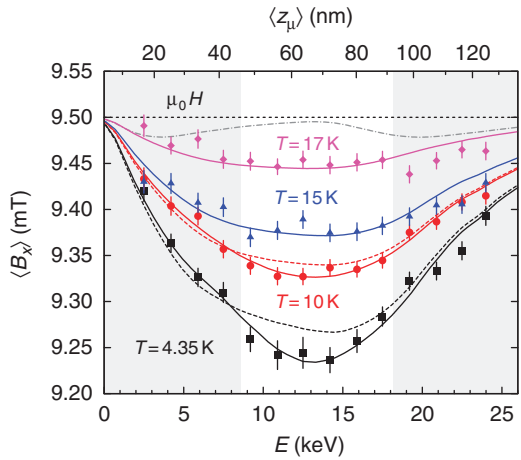

**Figure 6 | Depth profile of the local field at different temperatures.** The grey shaded areas indicate the top and bottom layers of the La$_{1.84}$Sr$_{0.16}$CuO$_4$ (46 nm)/La$_{1.94}$Sr$_{0.06}$CuO$_4$ (46 nm)/La$_{1.84}$Sr$_{0.16}$CuO$_4$ heterostructure. The horizontal dashed line shows the applied field of 9.5 mT. Points: measured average fields. The entire heterostructure excludes the magnetic flux like a superconductor: it shows the Meissner effect with the UD layer active in the screening. Note that in this geometry all the supercurrent must pass through the region of the heterostructure, which, taken as an isolated layer, would be in the normal state at $T > T_c'$. This functional form can only be observed if shielding supercurrents flow across (that is, along the $c$ axis) as well as in the $ab$ planes of the UD barrier. The lines are the fits using the London model described in the text. The fit takes into account the energy-dependent muon stopping profiles, which are also used to calculate the average stop depth $\langle z_\mu \rangle$ (upper scale). The grey dash-dotted line shows the field profile that would be expected (at $T = 4.35$ K) if the shielding current flow were restricted to the upper and lower superconducting electrodes. The dashed lines are obtained if one assumes that supercurrents in the barrier flow only in $c$ direction.

$\lambda$, and in the barrier, $\lambda'$. Specifically, the energy dependence of $\langle B_x \rangle$ is fitted with the mean field prediction of the model $\langle B_x \rangle = \int B_x(z) n(z) \mathrm{d}z$, where $n(z)$ is the simulated energy-dependent stopping distribution corrected for the fraction stopping in the magnetic region and $B_x(z)$ is the solution of the London model for three juxtaposed superconducting layers with magnetic penetration depth $\lambda$ and $\lambda'$ as the fitting parameters. The field profile is modelled by the form $B_x(z) = c_1 e^{-z/\lambda} + c_2 e^{z/\lambda}$ and $B_x(z) = c_3 e^{-z/\lambda} + c_4 e^{z/\lambda}$ for $z$ within the upper and bottom layer and $B_x(z) = c_5 e^{-z/\lambda'} + c_6 e^{z/\lambda'}$ for $z$ within the barrier. The coefficients $c_1$–$c_6$ are determined from the boundary conditions $B_x(0) = B_x(d) = \mu_0 H_{ext}$ and from the continuity of $B_x$ and of the vector potential at the interfaces ($A_y = -\mu_0 \lambda_{ab}^2 j_y = -\lambda_{ab}^2 \, \mathrm{d}B_x / \mathrm{d}z$, where $y$ is the direction parallel to the $ab$ planes).

In a clean homogeneous superconductor the penetration depth is directly related to the superfluid density as $\lambda = (m_{ab}/e^2 \mu_0 n_s)^{1/2}$, where $m_{ab}$ is the in-plane effective mass, $e$ the electron charge, $\mu_0$ the vacuum magnetic permeability and $n_s$ the superfluid density. In our inhomogeneous heterostructure, $\lambda$ and $\lambda'$ are the effective length scales indicative of the superfluid densities in the different layers. For example, at $T = 10$ K, we get $\lambda = 334 \pm 6$ nm and $\lambda' = 287 \pm 60$ nm, values comparable to the magnetic penetration depth in OP single crystals[17]. That these values and the corresponding effective superfluid densities are similar is intriguing and is possibly a further manifestation of the anomalous character of the proximity effect we observe here. However, above this temperature, $\lambda'$ has a much more pronounced temperature dependence than $\lambda$ (see Fig. 7b), indicating that the proximity-induced superfluid can be more easily

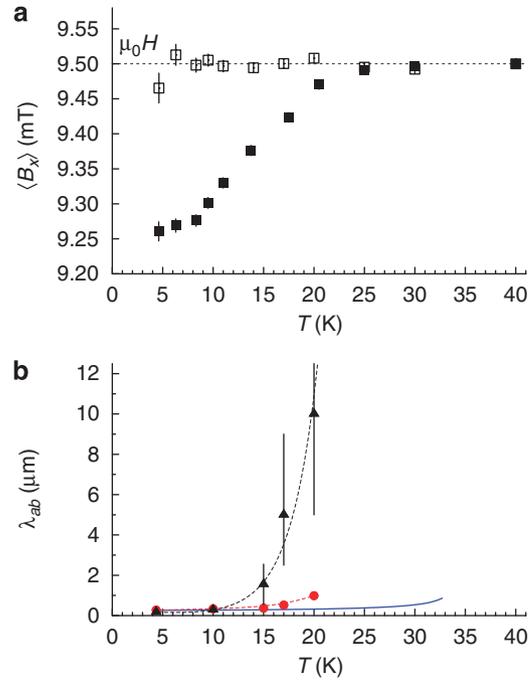

**Figure 7 | Temperature dependence. (a)** Field measured at the centre of the UD layer: as a single layer (open symbols) or as a barrier with thickness of 46 nm in the trilayer (filled symbols). In the latter case the average local field is diamagnetically shifted up to $T_{eff} \cong 22$ K. Above this temperature its value is within the experimental error equal to the applied field. No shift is observed for a single UD layer. **(b)** Temperature dependence of the magnetic penetration depths in the barrier (black triangles, $\lambda'$) and in the electrode layer (red circles, $\lambda$) compared with typical behaviour in optimally doped crystals (blue line) obtained from ref. 17. Error bars give the fit errors. The dashed lines are guides to the eyes. The divergent behaviour of $\lambda'$ close to 22 K indicates the disappearance of the long-range phase coherence in the barrier at that temperature. The temperature dependence indicates that the induced superfluid density in the barrier layer is more sensitive to thermal excitation than in a bulk superconductor.

suppressed by thermal excitations than the superfluid in intrinsically superconducting electrodes.

The observation of the Meissner effect and bulk diamagnetism at $T > T_c'$ in the UD La$_{1.94}$Sr$_{0.06}$CuO$_4$ barrier with thickness $d \gg \xi_c$ is the main result of this paper. The data presented here are more compelling and dramatic than the observation of unusual Josephson currents in similar structures. The 'bulk' diamagnetism we observe implies supercurrent flow without dissipation in response to an applied magnetic field, a hallmark of the Meissner effect. Moreover, because of the low $T_c'$ in the present experiment, the temperature window in which this effect is observed (up to over $4T_c'$) is much more extended than the range in which anomalous Josephson currents have been observed[7,8].

## Discussion

The conventional proximity theory in which the depth of penetration of Cooper pairs into a normal metal N is given by the induced coherence length $\xi_N$ cannot account for this observation[9,10]. In the usual situation, where the electron–electron interaction $V_N \rightarrow 0$ and $T_c' = 0$, one has $\xi_N \approx \hbar v_F / 2\pi k_B T$ in the clean limit (where $h \cdot 2\pi$ Planck constant, $v_F$ Fermi velocity and $k_B$ Boltzmann constant). A straightforward application of de Gennes expression taking into account that $\xi_N$ logarithmically diverges at $T = T_c' \neq 0$ may lead, even for cuprates, to values of the order of 10 nm (ref. 18). However, some caution has to be used when trying to apply the conventional proximity theory, which assumes a normal metal with $T_c' = 0$ and a well-defined isotropic Fermi velocity, to estimate the proximity length scales in cuprates. HTS are strongly correlated electron systems with a Fermi surface of complex shape and strongly anisotropic transport properties. Transport occurs more easily along the $ab$ planes, where it is band-like, than in the $c$ direction, where it is essentially incoherent hopping between planes with an effective velocity $v_c \ll v_f$. For $T \gg T_c'$, given that in UD cuprates the transport along $c$ axis is semiconducting, it is more appropriate to use the dirty-limit expression $\xi_N = (\hbar v_c l / 2\pi k_B T)^{1/2}$, where $l$ is the mean free path and $v_f/3$ is replaced by $v_c$. For $T > 8\,$K, this gives $\xi_N < 2.5\,$nm, much smaller than the barrier thickness $d = 46\,$nm.

To provide an enhanced length scale of the proximity effect, several models have been proposed[19–27]. One class of models postulate inhomogeneous barriers that contain superconducting islands embedded in a metallic matrix and forming a percolative network that can transmit supercurrent via Josephson coupling between the islands and also between the $CuO_2$ layers[19–22]. Another interesting theoretical proposal is the possibility of a new type of proximity effect between superconducting layers separated by an unconventional normal metal, such as a superconductor that has lost its phase rigidity due to phase fluctuations[23]. In this case, the well-defined homogeneous phase field of the S electrodes may quench the superconducting fluctuations present in the S′ barrier material, thus increasing the effective critical temperature of the barrier to some temperature $T_{eff}$ smaller than $T_c$ but well above $T_c'$[24,25]. An even more unconventional theory invokes a new mechanism of HTS—Bose–Einstein condensation of real-space pairs (bipolarons)[26]. Finally, it was also proposed that a topological Meissner effect, otherwise indistinguishable from the usual superconductor effect, could be generated by a chiral d-density wave in the pseudogap state even in the absence of superconductivity[28]. However, such an effect is predicted only for a magnetic field applied perpendicular to the $ab$ planes and should be absent in the present geometry.

To conclude, by performing local magnetic measurements in heterostructures digitally grown by molecular beam epitaxy, we observe a Meissner effect in a thick UD barrier layer well above its intrinsic critical temperature, $T_c'$. The induced superfluid density disappears at $T_{eff}$ where $T_c' \ll T_{eff} < T_c$. This result is not expected within the conventional proximity effect theory and constrains the theory of HTS and our understanding of the pseudogap.

The present work should be extended in several directions. Systematic measurements as a function of barrier thickness should be sensitive to the size of possible superconducting clusters. Furthermore, investigations as a function of doping up to the overdoped region, where more conventional mechanisms and length scales are expected, may help to elucidate the microscopic mechanism underlying this anomalous proximity effect.

## Methods

**μSR and low-energy μSR.** μSR makes use of polarized positive muons, which, thermalized in the sample, function as local magnetic probes in the host material. Conventional μSR uses energetic (~4 MeV) muons with stopping range and straggling on the sub-mm scale in a solid, so it is suitable for bulk studies.

LE-μSR uses 100% polarized positive muons of tunable energy in the keV range to study local (nanoscale) magnetic properties of thin films or heterostructures as a function of the muon penetration depth. These particles are produced by moderating an intense beam of energetic (~MeV) implanted muons in a condensed layer of solid $N_2$ or $Ar^{29}$. Suppression of the energy loss mechanisms responsible for the thermalization of muons results in a high probability for the emission of epithermal muons from the moderator with a kinetic energy of about 15 eV. The conservation of the spin polarization during the moderation process[30] permits the use of these epithermal muons as the source of a muon beam of variable kinetic energy between 0.5 and 30 keV and a narrow energy distribution. The corresponding implantation depth in solids ranges from a fraction of a nm to a few hundred nm. The stopping range profiles are calculated by the Monte Carlo program TRIM. SP, which treats the positive muon as a light proton[31,32].

**Time evolution of the muon spin polarization.** In the μSR experiment, few millions of spin-polarized muons are implanted one at a time into a sample, where they decay. The time evolution of the polarization of the muon ensemble $P(t)$ is followed via detection of the decay positron, which is emitted preferentially in the direction of the muon spin at the moment of decay. The number of positrons detected by a counter as a function of time after implantation reflects the time dependence of the muon spin polarization along the axis of observation defined by the detector:

$$N(t) = N_0 e^{-\frac{t}{\tau_\mu}} (1 + A_0 P(t)) + N_{Back} \qquad (2)$$

$A_0$ is the experimental decay asymmetry ($\cong 0.27$ in our spectrometer). $N_{Back}$ is a practically flat background of uncorrelated events.

The quantity $P(t)$ contains all the information about the interaction of the muon spin with its local magnetic or spin environment and therefore provides information on the host material. Each muon stopped in the lattice (typically in an interstitial site) experiences the net effect of external and local magnetic fields and precesses at the characteristic Larmor frequency $\omega_i = \gamma_\mu B$, where $\gamma_\mu/2\pi = 135.5$ MHz/T is the muon gyromagnetic ratio and $B$ is the total magnetic field at the muon site. μSR can measure local magnetic field distributions in the material. In a translationally invariant ferromagnet or antiferromagnet, a single precession frequency may be observed, without application of an external field (ZF measurement). If the muons experience different magnetic fields, $P(t)$ shows a distribution of precession frequencies with the corresponding width. If the field distribution is broad when averaged over the sample, only a fast relaxation of the muon polarization is observed.

In a paramagnetic environment, the polarization signal exhibits only slow relaxation and, if a magnetic field is applied perpendicular to the initial polarization direction (transverse field measurement, TF), a precession frequency corresponding to the external field is observed. In the case of different magnetic environments, $P(t)$ is the superposition of the corresponding signals. μSR is therefore able to detect and quantify different magnetic volume fractions in the sample.

## Acknowledgments

We thank M. Döbeli (ETH Zürich) for performing the Rutherford backscattering measurements and Z. Salman (PSI) for helping in the final phase of the LE-μSR measurements. The work at BNL was supported by the US Department of Energy, Basic Energy Sciences, Materials Sciences and Engineering Division.

## Author contributions

E.M. and I.B. proposed the present study. E.M. organized the research project with B.M.W. The Low-Energy μSR instrument at PSI was designed, tested and maintained by E.M., T.P. and A.S. B.M.W., E.M., A.S. and T.P. worked on the μSR data acquisition at PSI. B.M.W. and E.M. analysed the data. The films were synthesized using molecular beam epitaxy by G.L. and I.B. and characterized using RHEED, mutual inductance, XRD and atomic force microscopy. Additional characterization by XRD, reflectivity as well as resistivity measurements was performed by B.M.W. The text was drafted by E.M. with input from I.B. and B.M.W. All authors subsequently contributed to revisions of the text.

## Additional information

**Competing financial interests:** The authors declare no competing financial interests.

# Erratum: The Meissner effect in a strongly underdoped cuprate above its critical temperature

Elvezio Morenzoni, Bastian M. Wojek, Andreas Suter, Thomas Prokscha, Gennady Logvenov & Ivan Božović

This Article contains two typographical errors in the first and second paragraphs of the section entitled 'Diamagnetic response': '$\langle B_x \rangle \omega_{\rm L}/\gamma_\mu$' should be '$\langle B_x \rangle = \omega_{\rm L}/\gamma_\mu$' and '$\langle j_{ab} \rangle = \langle (1/\mu_0)\, dB_x/dz \rangle /\neq 0$' should be '$\langle j_{ab} \rangle = \langle (1/\mu_0)\, dB_x/dz \rangle \neq 0$'.